\documentstyle[epsfig,aps,twocolumn,prl,floats,tighten]{revtex}

\begin{document}

\title{Finite size effects and the order of a phase transition in 
fragmenting nuclear systems}

\author{J.M. Carmona$^a$, N. Michel$^b$, J. Richert$^b$
and P. Wagner$^c$\\[0.5em] 
$^a$ {\small Dipartimento di Fisica,}\\
{\small Via Buonarroti 2, Ed. B, 56127 Pisa (Italy)}\\[0.3em]
$^b$ {\small Laboratoire de Physique Th\'eorique,}
{\small Universit\'e Louis Pasteur,}\\
{\small 3, rue de l'Universit\'e, 67084 Strasbourg Cedex (France)}\\[0.3em]
$^c$ {\small Institut de Recherches Subatomiques,}\\
{\small BP28, 67037 Strasbourg Cedex 2 (France)}\\[0.3em]}

\maketitle

\begin{abstract}
We discuss the implications of finite size effects on the determination
of the order of a phase transition which may occur in infinite systems. We
introduce a specific model to which we apply different tests. They are
aimed to characterise the smoothed transition observed in a finite
system. We show that the microcanonical ensemble may be a useful framework
for the determination of the nature of such transitions.
\end{abstract}



\medskip


\medskip

Thermodynamics of finite systems.
Nuclear fragmentation.



\vskip 10 mm

Since the first attempt to use fragmentation experiments on 
nuclei~\cite{Pochodzalla,Serfling,Hauger,Ma} in
order to construct the caloric curve which links the temperature to the
excitation energy, many efforts have been made by different groups
in order to put the experimental results on firm grounds. Further
work is certainly necessary in order to eliminate as much as possible 
the remaining uncertainties which are present and difficult to get under
control. At the present stage it is not possible to see a clear-cut
sign for the existence of a phase transition and hence even more difficult
to get information about its order.

The approximate plateau observed in the first experimental measurements was 
interpreted as a sign for the presence of a first order liquid-gas phase 
transition~\cite{Pochodzalla}. This was in agreement with the
theoretical expectation that the equation of state should present the 
characteristics of a liquid-gas transition~\cite{Siemens}.
However, the latest analyses obtained by means
of peripheral collisions show a monotonously increasing curve and the onset
of a steep rise on the high energy side of the caloric
curve~\cite{Trautmann}.

This last result~\cite{Hauger,Kreutz,Schuttauf,Zheng} may be 
interpreted as a sign for a second order phase transition. It cannot be 
explained with the most commonly used
theoretical models showing a liquid-gas transition (Fisher droplet
model~\cite{Fisher1}, lattice gas model (LGM)~\cite{Campi}), which present
a single critical point in the phase diagram. 
This raises the question of the adequacy of the LGM which is the simplest
model describing a short-ranged interaction between classical particles 
inside a fixed volume.
A possible explanation of this fact has been proposed recently by
the authors of Ref.~\cite{Gulminelli}. They observed that finite size effects
could produce a scaling behaviour in fragment observables inside the 
coexistence region. These signals appear to be suppressed in the thermodynamic
limit.
If this scaling behaviour is related to the thermodynamic transition it
shows that a finite system can present misleading indications concerning
the order of the transition in the infinite system. However, these
indications are physical when one deals with small systems like fragmenting
nuclei. In this sense one can
speak about ``phase transitions in finite systems'' or ``crossovers'' 
(smoothed transitions). In the following we shall elaborate on this subject.

The LGM is usually formulated in the framework of the grand canonical  
ensemble. The basic variables of this model are the
temperature $T$ and the density $\rho$ of particles. In the grand canonical
formulation, the number of particles is not strictly fixed, but it is conserved
only in the mean. This model presents a first order transition for all values
of $\rho$ except for $\rho=0.5$, for which the phase transition is second 
order. In Ref.~\cite{Carmona} we considered a canonical framework for this 
model, the Ising model with fixed magnetization (IMFM). 
This \mbox{constraint} is important because it allows a direct
exploration of the coexistence region, which is forbidden in the grand
canonical formalism, but physically relevant because the system seen as
an ensemble of interacting particles can take
values of $(T,\rho)$ in this region. Since we are interested in 
finite crossovers, the ensemble in which the theoretical model is
formulated is important when we try to match it with experimental results.

 In the IMFM the number $A$ of particles which are
located on a lattice is strictly fixed in a fixed finite
volume $V=L^3$. The Hamiltonian reads
\begin{equation}
H_{\mathrm{IMFM}}=\sum_{i=1}^{A}\frac{p_i^2}{2m}+V_0\sum_{\langle ij\rangle}
\sigma_i\sigma_j,
\label{HamIMFM}
\end{equation}
where $V_0$ is a constant potential strength and $\{\sigma_i=\pm 1\}$.
The interaction acts between nearest neighbours.

 The Hamiltonian can be rewritten in terms of $s_i=(\sigma_i+1)/2$, the 
total number of particles is $\sum_i s_i=A$, and the density $\rho=A/V$ 
is fixed as a constraint so that the partition function 
in the canonical ensemble reads
\begin{equation}
Z=\sum_{\{\sigma\}} e^{-\beta H_{\mathrm{IMFM}}} \delta_{\sum_i s_i,A}\, ,
\label{ZIMFM}
\end{equation}
where $\beta=T^{-1}$ is the inverse temperature.

 In Ref.~\cite{Carmona} we worked out this model in $3d$ for 
finite systems with linear
dimensions ranging from $L=10$ to $L=48$. The energy sampling of simulation 
events and the 
determination of critical exponents relying on finite size scaling (FSS)
assumptions~\cite{Brezin,Fisher} lead to the conclusion that the system
experiences a second order transition at every value of the density $\rho$.

 This result was in contrast with the common understanding that the 
LGM has a first order transition line except for a single 
point located at $\rho=0.5$. 
The main argument to explain this difference was the following. If one looks 
at the phase diagram and lowers the temperature for fixed density
from above the transition line in the canonical
description of this model it is possible to cross this line and enter into the 
'coexistence' zone. In the grand canonical
formulation (LGM)~\cite{Carmona}, the states of this zone
are not accessible as equilibrium states. Indeed, in the IMFM  
the magnetisation $m=\sum_i \sigma_i$ which is equivalent to the density
$\rho$ is not discontinuous at the separation line as it should be in a 
first order transition.
 However, we could not firmly establish the nature of the transition in the
thermodynamic limit because of limitations imposed by numerics on the size of 
the largest system which could be generated. This is the reason for which 
we discuss different tests aimed to characterise the smoothed crossover.

\begin{figure}[tb]
\begin{center}
\begin{tabular}{cc}
\epsfig{figure=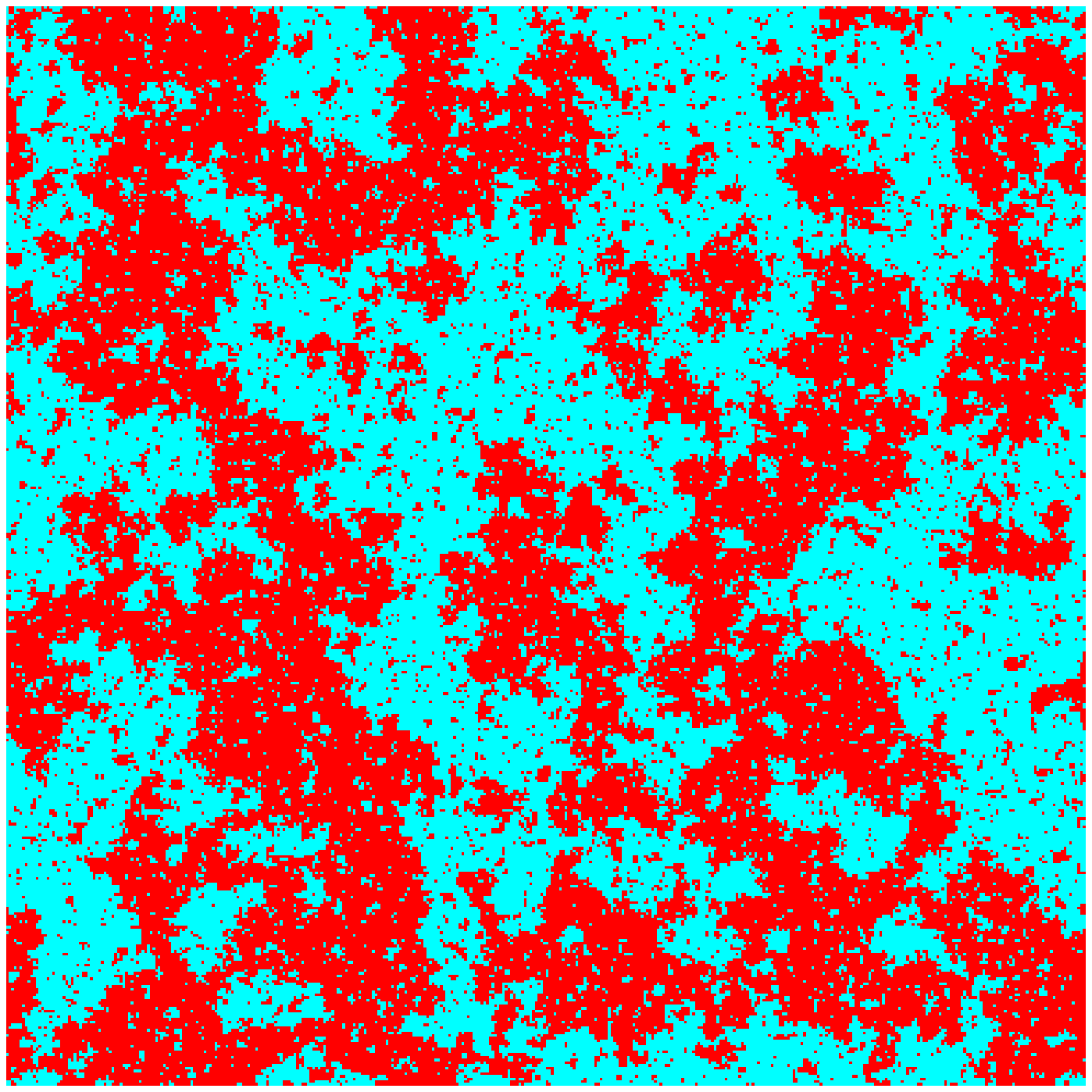,angle=0,width=40mm} &
\epsfig{figure=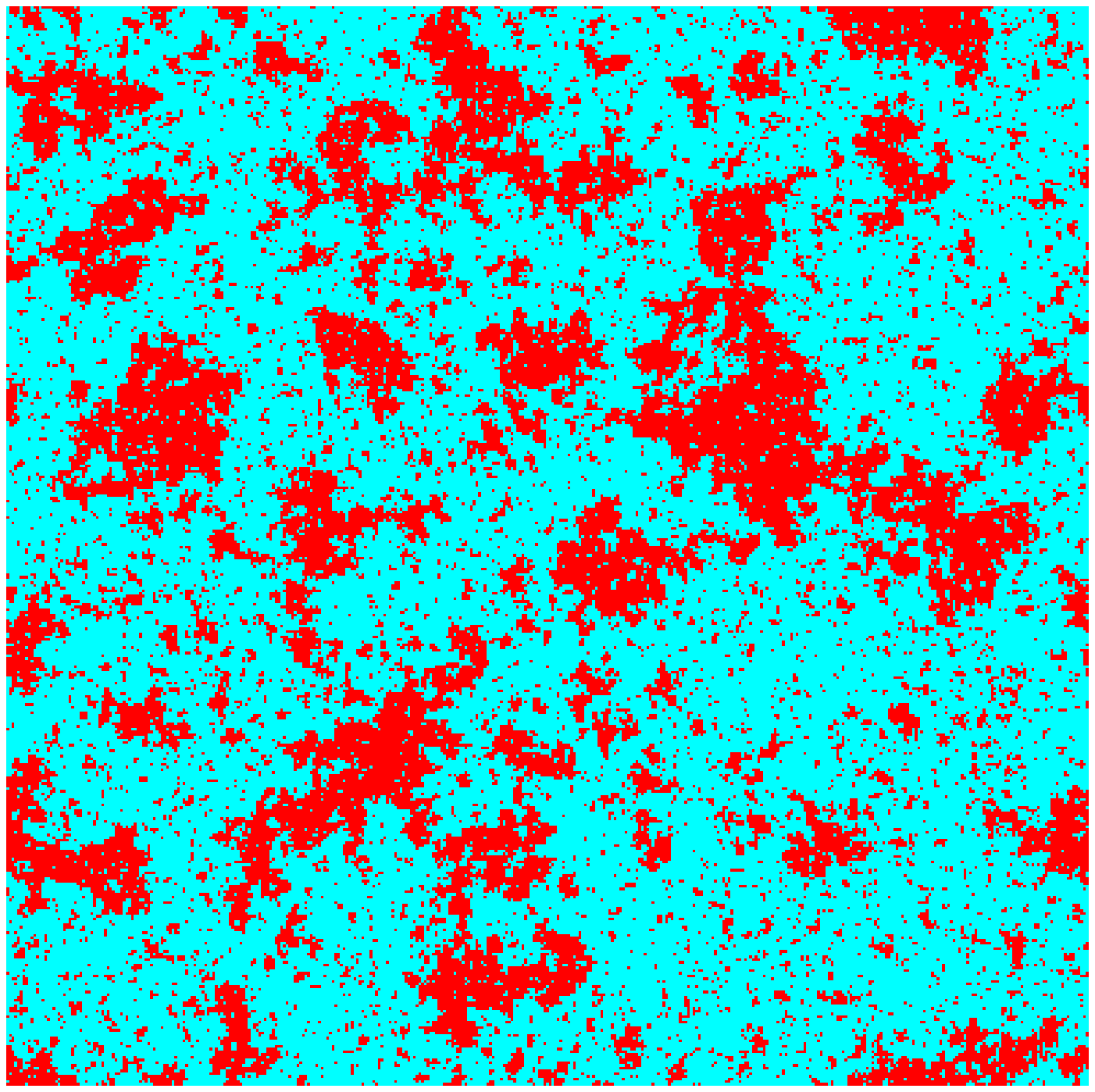,angle=0,width=40mm} \\
(a) & (b) \\
\end{tabular}
\end{center}
\caption{Distribution of particles in a $2d$ system calculated in the
framework of the canonical IMFM, at the phase separation line for
densities (a) $\rho=0.5$ and (b) $\rho=0.3$. The dark areas correspond
to particles, the grey areas to vacuum.}
\label{fig:homog}
\end{figure}

 The observed behaviour 
of the caloric curve and the specific heat cannot exclude a possible 
non-homogeneity in the 'coexistence' zone and the importance of surface 
effects. Hence our first test concerns  
the topology of the finite system in the vicinity of 
the line which separates the two phases.
 Fig.~\ref{fig:homog} shows two configurations generated in the framework of
the IMFM for a $2d$ system with linear size $L=400$. They correspond to two
values of the density, just below the transition line (low temperature side).
In the ordinary Ising model, the case $\rho=0.5$ corresponds 
to a second order phase transition, and $\rho=0.3$ to a first order 
transition. In the second case, the non-homogeneous configurations are those
which dominate in the thermodynamic limit below the transition. One expects
the system to be divided in two well-defined phases separated by a transition
domain of linear dimension $\xi$, where $\xi$ is the correlation length.
However, as one can see in Fig.~\ref{fig:homog}, in the IMFM the system looks 
rather homogeneous
in both cases and one expects a self-similar pattern at different scales
which is a qualitative indication of a continuous behaviour.  
At very low temperatures, the
system is made up of large and compact clusters and hence gets
non-homogeneous, but we checked that this happens for $\rho=0.5$ as well as 
for any other value of $\rho$.

\begin{figure}[tb]
\centerline{\epsfig{figure=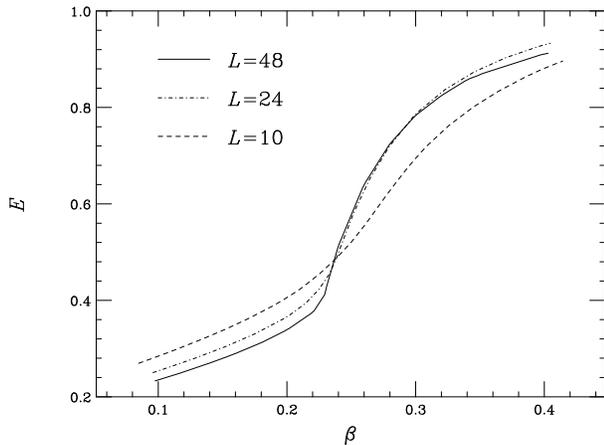,angle=90,width=80mm}}
\caption{Canonical caloric curves for $3d$ IMFM systems of linear sizes
$L=10, 24, 48$ with periodic boundary conditions and a density $\rho=0.3$.}
\label{fig:cc}
\end{figure}

 The next point concerns the caloric curve for the IMFM in $3d$. In the
thermodynamic
limit the caloric curve of a system which undergoes a first order transition
shows a plateau which signals the generation
of latent heat. This is why the first experimental caloric
curves~\cite{Pochodzalla}, 
apparently showing a plateau, were considered as a reminiscence of a 
liquid-gas phase transition present in nuclear matter. In the framework of
the IMFM, we constructed the caloric curve for different sizes of $3d$
systems. Fig.~\ref{fig:cc} shows these curves for $L=10, 24$ and 48 and a
density $\rho=0.3$. All three curves exhibit an inflection point and the slope
of the curve in the interval of energy where it increases rapidly gets 
steeper with increasing $L$. However, one observes only a very small change 
between $L=24$ and $L=48$ which seems to indicate that the asymptotic limit
is close. It is nevertheless clear that for $L=48$ there is no sign for the
appearance of a real plateau. Thus for this system one observes a 
behaviour of the caloric curve which indicates a continuous
transition between two homogeneous phases in the asymptotic limit. Such a
behaviour is in qualitative agreement with recent data 
analyses~\cite{Trautmann}.
Similarly to the FSS analysis, it again does not allow to establish what 
happens in the infinite system.

As a third test we introduce the microcanonical approach in order to 
compare the caloric curve with the same curve obtained in the framework of 
the canonical ensemble. It is well established that a closed thermalised
system with fixed energy and number of 
particles, which is a finite microcanonical ensemble
shows a caloric curve which is multivalued in energy for fixed 
temperature, i.e. shows
an oscillation (``S'' curve behaviour). As a consequence the specific heat
gets negative
for fixed volume over certain intervals of energy~\cite{Gross,Huller}.
This effect is due to the non-homogeneity of the system which characterizes
a first order transition. The surface energy of the clusters in the 
coexistence phase can be read from the area of the domains obtained 
by means of a Maxwell construction in the region of energy where the
two phases coexist~\cite{Gross}. The canonical caloric curve does not show 
this effect.

\begin{figure}[tb]
\centerline{\epsfig{figure=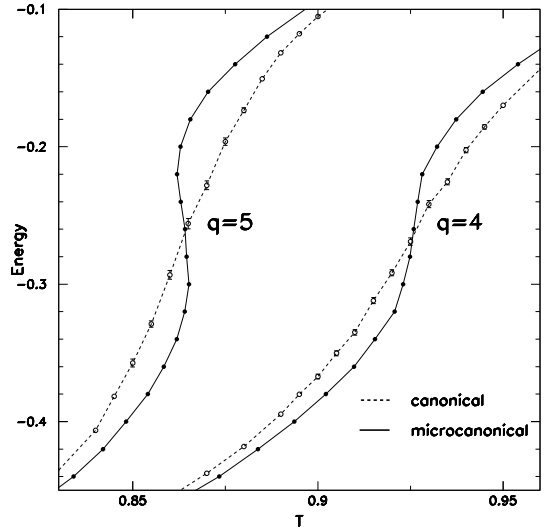,angle=0,width=80mm}}
\caption{Caloric curves for the $q=4$ and $q=5$ Potts model in a 
$16\times 16$ system. Black points (full line) are obtained from a 
microcanonical simulation, white points (dotted line)
from a canonical simulation. See comments in the text.} 
\label{fig:potts}
\end{figure}

 In order to test the existence or absence of this phenomenon in our model
we introduce a Metropolis Monte Carlo microcanonical algorithm~\cite{Ray}.
We apply it to the $q$-state Potts model~\cite{Potts} in 
two dimensions, which presents a second order transition for $q\leq 4$ and
a first order transition for $q\geq 5$ and is used here as a test model.
As expected, the system shows the two types of caloric curves, i.e. the 
characteristic  ``S'' behaviour of the caloric curve when the transition is 
first order ($q=5$), and a monotonous rise when the transition is second 
order ($q=4$), see Fig.~\ref{fig:potts}. This can be compared to
a canonical simulation of the Potts model which is also shown in the figure. 
The backbending is absent for $q=5$.

\begin{figure}[tb]
\begin{center}
\begin{tabular}{cc}
\epsfig{figure=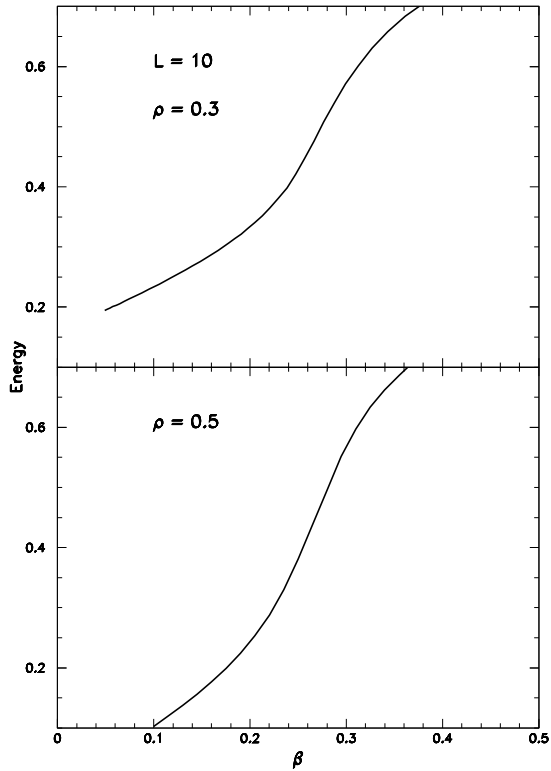,angle=0,width=120mm} \\
\end{tabular}
\end{center}
\caption{Caloric curve of the $d=3$ IMFM with $L=10$ at $\rho=0.3$
(upper part) and $\rho=0.5$ (lower part). Canonical and microcanonical
results are undiscernable. See comments in the text.}
\label{fig:imfm_l10}
\end{figure}

 In the case of the $3d$ IMFM, precise canonical and microcanonical
simulations lead to the results shown in Fig.~\ref{fig:imfm_l10}.
The figures correspond to densities of $\rho=0.3$ and $\rho=0.5$.
There exists no sign for the existence of multivaluedness of the energy
for fixed temperature which would indicate a first order transition.
We also observe from these figures that microcanonical and canonical results
are already the same for a still relatively small lattice volume ($L=10$).
This result supports the previous indications of a continous transition in
the thermodynamic limit.

\begin{figure}[tb]
\begin{center}
\begin{tabular}{cc}
\epsfig{figure=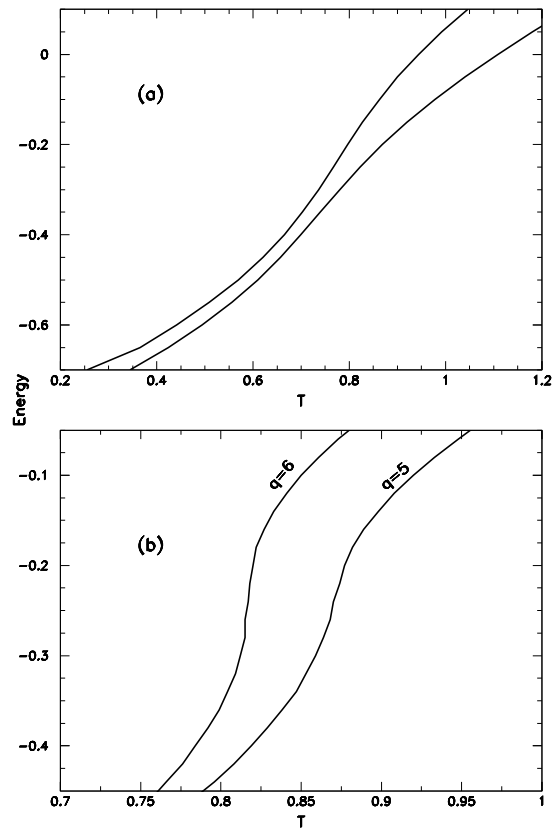,angle=0,width=120mm} \\
\end{tabular}
\end{center}
\caption{Constrained $d=2$ Potts model caloric curves in the framework of 
the microcanonical ensemble, for  $L=16$. 
(a) $q=5$ and fixed spin population. Upper curve : $N(p)$ fixed to values
51,51,51,51,52 for $p=0,1,2,3,4$ respectively. Lower curve :
$N(p)$ fixed to values 156,25,25,25,25 for $p=0,1,2,3,4$ respectively.
(b) $q=5$ and 6 with the constraint $N(0)=17$. Here the constraint is less 
stringent since only one type of spin directions has been fixed.
See comments in the text.}
\label{fig:constraint}
\end{figure}

 These results raise the general question of the role played by constraints 
on the order of the transition. 
In order to investigate this point further 
we consider the Potts model for $q\geq 5$ and introduce different 
constraints. In practice we fix the number $N(p)$ of spins in a given
direction $(p=0,\ldots,q-1)$ to a given value and generate the caloric 
curve for the systems in the framework of the microcanonical ensemble. Some 
results are shown in Fig.~\ref{fig:constraint}. One sees that the curves 
no longer correspond to a multivalued temperature $T$ as a function of 
energy but show a smooth increase, with the presence of an inflection point 
as expected for a second order transition when one goes to the 
thermodynamic limit. This result goes the same direction as the result 
obtained from the IMFM and confirms indeed the fact that constraints 
may have drastic effects on the order of the transition.

In summary, different tests show that the IMFM, in its
microcanonical or canonical formulation, exhibits the 
characteristic features of an homogeneous system 
which will undergo a smooth transition
from one phase to the other in the thermodynamic limit.
It may be surprising to find such a behaviour in the framework of the LGM,
but the constraints imposed on the model seem to produce this effect.
The strongest argument pointing
towards this conclusion concerns the microcanonical test which indicates that
the expected effect characterizing a discontinuous transition on the caloric
curve is not seen. This is also in agreement with the finite size scaling 
analysis carried out in Ref.~\cite{Carmona}. 
It shows that microcanonical calculations 
may be of help in the characterisation of the phase crossover
in finite systems.
As already stated above, we however cannot conclude that the phase transition
is indeed necessarily of second order in the thermodynamic limit. Hence
it may also be difficult to decide about the order of a phase transition in
infinite nuclear matter from experimental results extracted from finite nuclei
collisions.

 The sharpest experimental test would correspond to measurements in which 
energy and number of particles are strictly fixed and hence the fragmenting 
system could be assimilated to a microcanonical ensemble. Its caloric curve 
would then be able to reveal not only the characteristics of the crossover
in the finite system but also the nature of the thermodynamic transition. 
Very recently it was claimed that the analysis of fragmentation events 
show negative values for the specific heat~\cite{D'Agostino}. 
The confirmation of these
results would be a proof for the presence of a first order phase transition
and confirm the predictions of models which where proposed in the near past
\cite{Campi,Pan,Gulminelli,Borg}.

 We thank A. Taranc\'on for his help and interest in this work, and V\'{\i}ctor
Laliena and Ettore Vicari for useful discussions. Work partially 
supported by the EC TMR program ERBFMRX-CT97-0122.

\end{document}